\documentclass[article,twocolumn,showpacs,preprintnumbers,amsmath,amssymb]{revtex4-1}
\usepackage{graphicx}
\usepackage{amsmath}
\usepackage{booktabs}
\usepackage{epsfig}
\usepackage{subfigure}
\usepackage{amssymb,amsfonts,amsmath}

\usepackage{float}

\makeatletter
\renewcommand{\section}{\@startsection{section}{1}{0mm}
  {-\baselineskip}{0.5\baselineskip}{\bf\leftline}}
\renewcommand{\subsection}{\@startsection{subsection}{1}{0mm}
  {-\baselineskip}{0.5\baselineskip}{\bf\leftline}}

\makeatother

\begin{document}


\title{Extortion under Uncertainty: Zero-Determinant Strategies in Noisy Games}

\author{Dong Hao $^{a)}$}

\author{Zhihai Rong $^{a)}$}

\author{Tao Zhou $^{a,b)}$}
\email{zhutou@ustc.edu}

\affiliation{
$^{a)}$ CompleX Lab, Web Sciences Center, University of Electronic Science and Technology of China, Chengdu 611731, P.R. China\\
$^{b)}$ Big Data Research Center, University of Electronic Science and Technology of China, Chengdu 611731, P.R. China
}

\begin{abstract}

%
Repeated game theory has been one of the most prevailing tools for understanding the long-run relationships, which are footstones in building human society. Recent works have revealed a new set of ``zero-determinant (ZD)'' strategies, which is an important advance in repeated games. A ZD strategy player can exert a unilaterally control on two players' payoffs. In particular he can deterministically set the opponent's payoff, or enforce an unfair linear relationship between the players' payoffs, thereby always seizing an advantageous share of payoffs. One of the limitations of the original ZD strategy, however, is that it does not capture the notion of robustness when the game is subjected to stochastic errors. In this paper, we propose a general model of ZD strategies for noisy repeated games, and find that ZD strategies have high robustness against errors. We further derive the pinning strategy under noise, by which the ZD strategy player coercively set the opponent's expected payoff to his desired level, although his payoff control ability declines with the increase of noise strength. Due to the uncertainty caused by noise, the ZD strategy player cannot secure
his payoff to be higher than the opponent's, which implies \emph{strong extortions} do not exist even under low noise. While we show that the ZD strategy player can still establish a novel kind of extortions, named \emph{weak extortions}, where any increase of his own payoff always exceeds that of the opponent's by a fixed percentage, and the conditions under which the weak extortions can be realized are more stringent as the noise becomes stronger.

\end{abstract}


\maketitle

\section{Introduction}
Repeated games have been representative to explore the agents' long-run relationships, which help us in understanding how cooperation and competition might arise among agents with selfish objectives. Extensive literatures have by now utilized repeated games as a basic component to analyze economic behaviors, political sciences, evolutionary dynamics as well as multi-agent systems \cite{M&LBOOK}. It has been commonly accepted that in such games there is no simple ultimatum strategy whereby one player can simply occupy an unfair share of the payoffs. However, Press and Dyson's discovery of ``zero-determinant (ZD)'' strategies illuminates a new starting point \cite{Press2012}. They show that in iterated Prisoner's Dilemma, it is possible for a player (named ZD strategy player, ZD player for short) to unilaterally enforce a linear relationship between his and the opponent's payoff, thereby deterministically setting the expected payoff of the opponent to a fixed value or ensuring that, when the opponent tries to increase his payoff, he will always increase the ZD player's payoff even more. The discovery of ZD strategies is a milestone along the way to fundamentally understand how different strategies correlate with each other and what are the underlying norms of social interactions \cite{Stewart2012,Hayes2013}. It provides us with a powerful but succinct framework for motivating and sustaining the cooperation required for any society, as well as for controlling the damages done by the unscrupulous or mischievous agents.

ZD strategies have thus attracted considerable attentions and been incorporated successfully into a wide array of researches, ranging from theoretical game researches to real-world experimental studies \cite{Hao2014}. Among the subsequent researches, Roemheld generalizes ZD strategies for all symmetric bimatrix games as well as for the Battle of the Sexes, which is the most common example for the asymmetric games \cite{Roemheld2013}.
Akin explores a broader space of strategies by extending Press-Dyson theorem, and obtains the cooperation-enforcing good strategies \cite{Akin2012}. Thereafter, Stewart and Plotkin, as well as Hilbe et al. identify the intersection of ZD strategies and good strategies, named generous ZD strategies, which not only control the payoffs, but also cooperate with others and forgive defecting opponents, leading the game towards a win-win situation \cite{Stewart2013,Hilbe2013Plos}.  Chen and Zinger analyze the robustness of ZD strategies against evolutionary players and prove that there always exist evolutionary paths for ZD player to obtain the maximum payoff \cite{Chen2014}. Press and Dyson's work can be further generalized to multi-player ZD strategies for investigating various social dilemmas, new features and constrains related to participant number and payoff structure have been revealed and the impact of ZD alliance in multi-player games has been studied \cite{Pan2014,Hilbe2014}. Furthermore, there are also extensive literatures investigating the significance of ZD strategies in evolutionary game theory and in social networks \cite{Stewart2013,Hilbe2014,Hilbe2014-2,Hilbe2013,Hilbe2013Plos,Adami2013,Szolnoki2014,Wu2014,Szolnoki2014-2}. Although initially the evolutionary instability was found for extortion strategies \cite{Adami2013}, later it is proved that the generous strategies finally dominate in population and are stable in an evolutionary sense \cite{Stewart2013,Hilbe2013}. The above theoretical studies also have been implemented in real-world social experiments, it is confirmed that extorting others has limited prospects, and in the long run, generosity is more profitable \cite{Hilbe2014-2}.

By now, how the ZD strategies perform in realistic noisy games is still an open problem. As in Stewart and Plotkin's commentary to Press and Dyson's work \cite{Stewart2012}, one of the key questions is: how does ZD strategies fare in iterated games in the presence of noise?
Since stochastic perturbations due to observation errors, action mistakes, biological mutations and other chance events are common and inevitable in reality, it is of great importance to extensively investigate the strategies and solutions in games theory at the presence of noise. However, the majority of known results on game theory \cite{Kandori2002}, as well as those related to ZD strategies \cite{Hao2014}, are obtained in a perfect environment without any noise. Actually, the analyses of noisy repeated game have been long-standing challenges and are at the cutting edge of researches on game theory and social interactions \cite{Kandori2002, Fudenberg1990, Fudenberg2012, Nowak1995,Sekiguchi1997,Barlo2009,Mailath2002,Mailath2011}. The errors in noisy repeated games usually fall into two categories \cite{Nowak1995}. The first kind is that players' actions are often observed with errors, which can be called the \emph{perception errors}: someone who claims they worked hard, or that they were too busy to give a help, may or may not be telling the truth; similarly, awkward results sometimes accidentally come after good behaviors \cite{Fudenberg2012}. The second kind is that players may wrongly take an action. This is categorized into \emph{implementation errors} (or \emph{action errors} in the literature): one player has intended action, but may accidentally chooses another action due to interferences, this is also described by the well-known notion ``trembling hands'' \cite{Fudenberg1990}.

To explore the noisy games, a virgin land for ZD strategies, we propose a general framework of ZD strategies in noisy repeated games, and show the implementable for a unilateral payoff control. Since repeated games with perception errors are the most stringent case \cite{Kandori2002,M&LBOOK}, our analysis focus primarily on this scenario, and it can be easily extended to repeated games with implementation errors.
It is found that, ZD strategies present strong robustness against noise. Even in environments with perception or implementation errors, a player can still enforce a linear relationship between the two players' payoffs. Under noisy repeated games, we classify the ZD strategies into three subsets, (i) pinning strategies, (ii) weak extortion strategies and (iii) strong extortion strategies. Following the pinning strategy, the ZD player can unilaterally set the opponent's payoff to his desirable level, although the difficulty for realizing such payoff control increases as the noise becomes stronger. Furthermore, we prove that since the noise brings uncertainty and risk to the ZD player, he cannot perfectly secure his payoff to be always greater than that of the opponent. That is to say, \emph{strong extortion} strategies do not exist even when the noise strength is low. Nevertheless, the ZD player can still ensure that his own increase of payoff always exceeds that of the opponent by a fixed percentage, such that as long as the opponent tries to improve his payoff, he will improve the ZD player's payoff even more, and the opponent can only maximize his payoff by fully cooperating, then both players' payoffs are maximized but the ZD player outperforms the opponent. We call such strategy as the \emph{weak extortion} strategy. The weak extortion strategy is close to the strong extortion strategy and the difference between them is caused by the noise structure. Our study implies that noises expose the ZD player to uncertainty and risk of losing, while the mischievous manipulation and the unusual control still stubbornly persist. The results of our study can be utilized both to propose a generalized framework for the ZD strategy
paradigm that has characterized much of the recent literatures and to provide a unilateral payoff control scheme for a larger class of noisy repeated games where payoff control is of great significance but has barely been studied.


\section{Noisy Repeated Game}
Consider two players engaged in an iterated prisoner's dilemma (IPD) game. In each stage, each player $i\in\{X,Y\}$ takes an action $a_i\in\{C,D\}$. Each player cannot directly see what action the opponent has taken, but only observes a private signal ${ \omega_i}\in\{g,b\}$, where $g$ and $b$ denote good and bad signals, respectively. Each player's signal $\omega_i$ is a stochastic variable, affected not only by the two players' actions but also by the noises (random errors) from the environment. Given the actions, every possible signal profile occurs with a positive probability $\pi({\bf \omega} | {\bf a})$, where ${\bf \omega}=\{ \omega_X, \omega_Y\}$ and ${\bf a}=\{a_X, a_Y\}$ are the observed signal profile and the action profile, respectively. In each stage, if player Y chooses $a_Y=C$ (or $a_Y=D$) but X observes $\omega _X=b$ (or $\omega _X=g$), it means an \emph{error} occurs. Denote $\tau$ the commonly known probability that neither player has an error, $\varepsilon$ the probability that an error occurs to only one player, and $r$ the probability that an error occurs to both players, obviously, $\tau+2\varepsilon+r=1$. Normally, the values
follow the order $\tau>\varepsilon>r>0$ which means the observations of players are more likely to be correct. For example, if both players take action $C$, then $\pi \left( {g ,g |CC} \right)=\tau$,
$\pi \left( {g ,b |CC} \right)=\pi \left( {b ,g |CC} \right)=\varepsilon$, and
$\pi \left( {b ,b |CC} \right)=r$.
The following tables summarize the signal
distributions under all action profiles. Based on the action and privately observed signal, for a player X, his \emph{private outcomes} in each stage game is a tuple $(a_X, \omega_X ) \in \{Cg, Cb, Dg, Db\}$. Note that this is different from games without noise, where both players' outcomes are identical and are just action profiles.
{\tiny
\begin{table}[h]
   \caption{\label{tab:signaldis}Signal distributions for different action profiles}
    \begin{tabular}{|c|c|c|}\hline
      $CC$ & $\omega_Y=g$ & $\omega_Y=b$ \\\hline
      $\omega_X=g$ & $\tau$ & $\varepsilon$ \\\hline
      $\omega_X=b$ & $\varepsilon$ & $r$ \\\hline
    \end{tabular}
    \label{tab:general-pd-cc-sigdis}
    \begin{tabular}{|c|c|c|}\hline
      $CD$ & $\omega_Y=g$ & $\omega_Y=b$ \\\hline
      $\omega_X=g$ & $\varepsilon$ & $r$ \\\hline
      $\omega_X=b$ & $\tau$ & $\varepsilon$ \\\hline
    \end{tabular}\\

    \begin{tabular}{|c|c|c|}\hline
      $DC$ & $\omega_Y=g$ & $\omega_Y=b$ \\\hline
      $\omega_X=g$ & $\varepsilon$ & $\tau$ \\\hline
      $\omega_X=b$ &  $r$ & $\varepsilon$ \\\hline
    \end{tabular}
    \begin{tabular}{|c|c|c|}\hline
      $DD$ & $\omega_Y=g$ & $\omega_Y=b$ \\\hline
      $\omega_X=g$ & $r$ & $\varepsilon$ \\\hline
      $\omega_X=b$ &  $\varepsilon$& $\tau$ \\\hline
    \end{tabular}
\end{table}
}

Since the stochastic changes of the environment as well as the opponent's action is jointly involved in the signals, the realized stage payoff for each player depends only on the action he chose and the signal he received, denoted as $u_i(a_i,\omega_i)$ \cite{M&LBOOK,Kandori2002,Sekiguchi1997}. Assume that the realized stage payoff follows the prisoner's dilemma, such that
${u_i}\left( {C,g} \right) = 1$,
${u_i}\left( {C,b} \right) =  - L$,
${u_i}\left( {D,g} \right) = 1 + G$, and ${u_i}\left( {D,b} \right)  = 0$, where $L$ and $G$ are positive variables. According to the general framework in \cite{Sekiguchi1997},  in each stage, player $i$'s expected payoff when two players have an action profile ${\bf{a}}$ is derived as
\begin{eqnarray}\label{expected_stage_payoff}
{f_i}\left( {\bf{a}} \right) = \sum\limits_{\bf \omega}  {{u_i}\left( {{a_i},{\omega _i}} \right)\pi \left( {{\bf \omega} |{\bf{a}}} \right)},
\end{eqnarray}
such that ${f_i\left( {\bf{a}} \right)}$ is the expected value over all possible signals, conditioning on the two players' actions.
The expected payoffs under different action profiles $CC$, $CD$, $DC$ and $DD$ are denoted as $R_E$, $S_E$, $T_E$ and $P_E$, which can be respectively calculated according to Eq. (\ref{expected_stage_payoff}), as $R_E=1-(L+1)(\varepsilon+r)$, $S_E=-L+(1+L)(\varepsilon+r)$, $T_E=(1+G)(1-\varepsilon-r)$ and $P_E=(1+G)(\varepsilon+r)$. Then player X's expected stage payoff vector is denoted as ${\bf U_X}=(R_E,S_E,T_E,P_E)$ and player Y's is denoted as  ${\bf U_Y}=(R_E,T_E,S_E,P_E)$. 

\begin{figure}[h]
  \subfigure{\label{fig:gull}\includegraphics[width=0.46\textwidth]{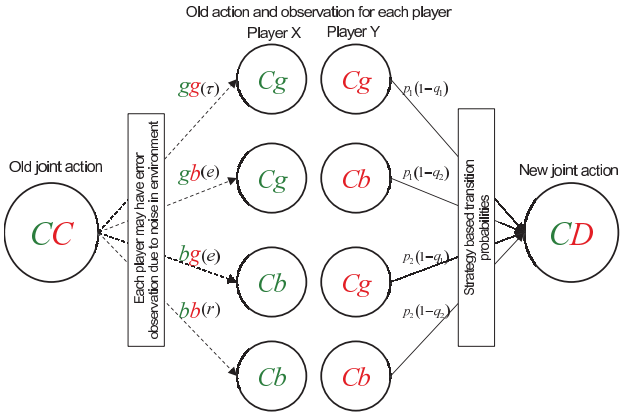}}
  \caption{Illustration of the transition from state (joint action) $CC$ to $CD$. The green color shows the real action and observation of player X while the red color depicts that of player Y. The big nodes denote the action profile, which is the real states of the game. The small nodes denote the combination of one player's  action and observation, which are one player's private outcomes.}
  \label{figure1:feasible_region_and_correponding_payoffs_of_pinning_strategies}
\end{figure}

We concentrate on the memory-one strategies where each player sets his strategy only according to the single previous outcome \cite{Press2012,Barlo2009,Murdock1962}.  
Denote the probabilities that player X will cooperate under his previous outcomes $Cg, Cb, Dg$ and $Db$ as $p_1, p_2, p_3$ and $p_4$ and the
probabilities that Y will cooperate under her previous outcomes $Cg, Cb, Dg$ and $Db$ are $q_1, q_2, q_3$ and $q_4$. The joint actions of the two players are the \emph{states} of the game, and the two players' probabilistic strategies as well as the noise structure jointly determine the transition rule of the states. Note that the observation errors only changes the transition probabilities, but never changes the real state space of the game, which is still $\{ CC, CD, DC, DD \}$. For example,
if the old state is $CC$, the probability that the state transits to
a new joint state $CD$ will be:
$
\tau p_1 \left( {1 - q_1 } \right) + \varepsilon p_1 \left( {1 - q_2
} \right) + \varepsilon p_2 \left( {1 - q_1 } \right) + rp_2 \left(
{1 - q_2 } \right) ,
$
where $\tau p_1 \left( {1 - q_1 } \right)$ is the probability that
both players observe correct signals and player X
takes action $C$ while player Y takes action $D$ in the new stage;
$\varepsilon p_1 \left( {1 - q_2
} \right)$ and $ \varepsilon p_2 \left( {1 - q_1 } \right)$ are the probabilities one player has an observation error and player X
takes $C$ and player Y takes $D$; and $rp_2 \left(
{1 - q_2 } \right)$ is the probability that both players have observation errors and player X
takes $C$ and player Y takes $D$. The derivation of the transition probability from state $CC$ to state $CD$ is depicted in Figure \ref{figure1:feasible_region_and_correponding_payoffs_of_pinning_strategies}.

\begin{figure}[ht]
  \subfigure{\label{fig:gull}\includegraphics[width=0.49\textwidth]{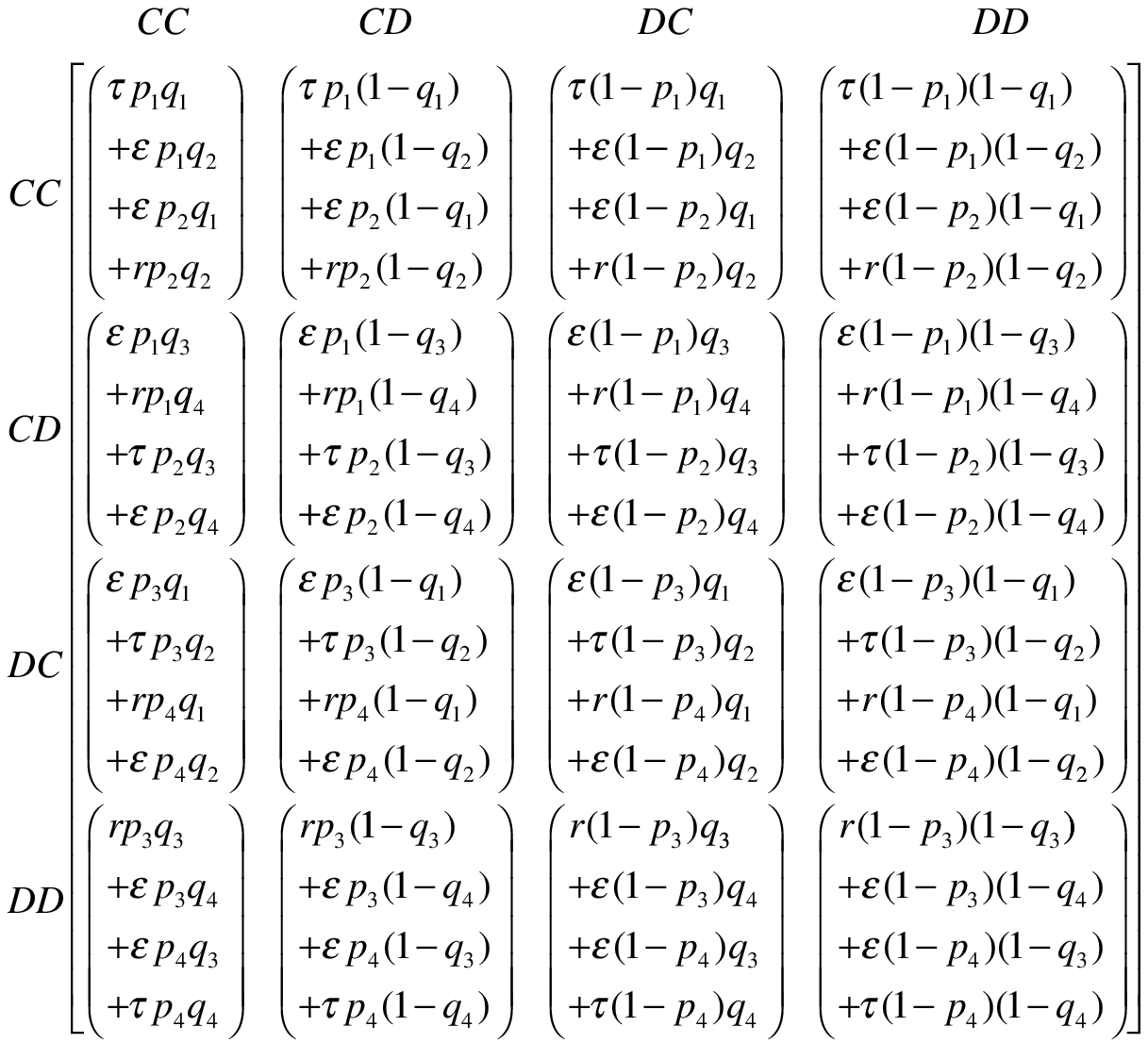}}
  \caption{Transition matrix of noisy repeated game.}
  \label{figure1:transition_matrix}
\end{figure}

This figure illustrates that the noise decomposes the state $CC$ into four combinations of private outcomes, namely $(Cg, Cg)$, $(Cg,Cb)$, $(Cb,Cg)$ and $(Cb,Cb)$. Following the same way, the state transition matrix $\bf{M}$ of the noisy repeated game is thus calculated as the matrix in Figure \ref{figure1:transition_matrix}. We can see from this transition matrix, although it becomes more complex, it is still a stochastic matrix.

\section{ZD Strategies under Noise}
Let $ {\bf{u}}^t $ be the probability
distribution over the game's state space $\{ CC,CD,DC,DD \}$ at stage $t$. The probability distributions follow
the transition rule such that
$
{\bf{u}}^{t + 1}  = {\bf{u}}^t  \times \bf{M}.
$
The \emph{stationary distribution} for ${\bf{M}}$ is a vector $\bf v$ such that
$
{\bf{v}^T}\bf{M}= {\bf{v}^T}.
$
Introducing ${\bf M'}={\bf M}-{\bf I}$ into the above equation yields ${\bf v}^T{\bf M'}={\bf 0}$. According to Cramer's rule, for any matrix ${\bf M'}$ and its adjugate matrix $Adj({\bf M'})$, the equation $Adj({\bf M'}){\bf M'}=\bf 0$ holds. Therefore from these two equations we know that every row of $Adj({\bf M'})$ is proportional to the stationary distribution vector ${\bf v}$. Changing the last column of ${\bf M'}$ into X's stage payoff vector $(R_E,S_E,T_E,P_E)$, we get a new matrix ${\bf{\tilde M}}$. Then using Laplace expansion on the last column of ${\bf{\tilde M}}$, we have $
\det ( {\bf{\tilde M}} ) =  R_E  \cdot N_1 + S_E  \cdot N_2  + T_E  \cdot N_3  + P_E  \cdot N_4.
$
The variables $N_1$, $N_2$, $N_3$ and $N_4$ are just the minors corresponding to $R_E$, $S_E$, $T_E$ and $P_E$ in the last column of ${\bf{\tilde M}}$, respectively. The fourth row of $Adj({\bf{\tilde M}})$ is calculated from the first three columns of ${\bf{\tilde M}}$ and is always proportional to ${\bf v}$. Therefore X's expected payoff can be calculated by using $\det ( {\bf{\tilde M}} )$. Adding the first column into the second and the third columns gives us a new form of this determinant as in Eq. (\ref{det_for_X}).
{\small
\begin{widetext}
\begin{eqnarray}
\begin{array}{l}\label{det_for_X}
\begin{array}{l}
 \det ( {\bf{\tilde M}} )
  = \left| {\begin{array}{*{20}c}
    \cdots  & {\kern 18pt}{   (\tau  + \varepsilon )p_1  + (r + \varepsilon )p_2- 1  } & {\kern 18pt}{   (\tau  + \varepsilon )q_1  + (r + \varepsilon )q_2- 1  } & {\kern 18pt}{R_E }  \\
    \cdots  & {\kern 18pt}{  (r + \varepsilon )p_1  + (\tau  + \varepsilon )p_2 - 1 } & {\kern 18pt}{(\tau  + \varepsilon )q_3  + (r + \varepsilon )q_4 } & {\kern 18pt}{S_E}  \\
    \cdots  & {\kern 18pt}{(\tau  + \varepsilon )p_3  + (r + \varepsilon )p_4 } & {\kern 18pt}{   (r + \varepsilon )q_1  + (\tau  + \varepsilon )q_2- 1  } & {\kern 18pt}{T_E}  \\
    \cdots  & {\kern 18pt}{(r + \varepsilon )p_3  + (\tau  + \varepsilon )p_4 } & {\kern 18pt}{(r + \varepsilon )q_3  + (\tau  + \varepsilon )q_4 } & {\kern 18pt}{P_E}  \\
\end{array}} \right| \\
 \end{array}.
\end{array}
\end{eqnarray}
\end{widetext}
}
In this determinant, the first columns is omitted because we only need to analyze the relationship between the second column and the fourth column. What's more important, we can see that in this determinant, the second column is solely controlled by X and
the third column is solely controlled by Y. Record this new format of determinant as $D\left( {{\bf{p,q,}}{{\bf{U}}_{\bf{X}}}} \right)$. Then, player X's normalized payoff score under stationary state is derived as

\begin{eqnarray}
{s_X} = \frac{{{\bf{v}} \cdot {{\bf{U}}_{{X}}}}}{{{\bf{v}} \cdot {\bf{1}}}} = \frac{{D\left( {{\bf{p,q,}}{{\bf{U}}_{{X}}}} \right)}}{{D\left( {{\bf{p,q,1}}} \right)}}.
\end{eqnarray}
Similarly, replacing the last column of $\det ( {\bf{\tilde M}} )$ by player Y's stage expected payoff vector, player Y's normalized payoff score is
\begin{eqnarray}
{s_Y} = \frac{{{\bf{v}} \cdot {{\bf{U}}_{{Y}}}}}{{{\bf{v}} \cdot {\bf{1}}}} = \frac{{D\left( {{\bf{p,q,}}{{\bf{U}}_{{Y}}}} \right)}}{{D\left( {{\bf{p,q,1}}} \right)}}.
\end{eqnarray}A linear combination of these two scores with coefficients $\alpha$, $\beta$ and $\gamma$ gives us
\begin{eqnarray}\label{mischivous}
\alpha {s_X} + \beta {s_Y} + \gamma  = \frac{{D\left( {{\bf{p,q,}}{\alpha{\bf U}_X+\beta{\bf U}_Y+\gamma{\bf U}_Z}} \right)}}{ D\left({\bf p},{\bf q},{\bf 1}\right) }.
\end{eqnarray}

If player X can set his strategy $\bf{p}$ delicately and make the second column of this determinant satisfy $ {\bf{\tilde
p}} = \alpha{\bf U}_X+\beta{\bf U}_Y+\gamma{\bf U}_Z $, then the determinant's value $ D\left({\bf p},{\bf q},{\alpha{\bf U}_X+\beta{\bf U}_Y+\gamma{\bf U}_Z}\right) =0$, which indicates that X can
unilaterally establish a linear relationship between X's and
Y's payoff scores, such that: $\alpha s_X + \beta s_Y  + \gamma  = 0$. Such linear relationship also requires a feasible solution to the following linear equation set:
\begin{eqnarray}
\begin{array}{l}\label{equationset}
\left\{ \begin{array}{l}
 (\tau  + \varepsilon )p_1  + (\varepsilon  + r)p_2   - 1 = \alpha R_E + \beta R_E + \gamma,  \\
  (\varepsilon  + r)p_1  + (\tau  + \varepsilon )p_2  - 1  = \alpha S_E + \beta T_E + \gamma,  \\
 (\tau  + \varepsilon )p_3  + (r + \varepsilon )p_4  = \alpha T_E + \beta S_E + \gamma,  \\
 (r + \varepsilon )p_3  + (\tau  + \varepsilon )p_4  = \alpha P_E + \beta P_E + \gamma  .\\
 \end{array} \right.
\end{array}
\end{eqnarray}
If this system of linear equations has feasible solutions, then it will be possible for player X to adjust $p_1, p_2, p_3$ and $p_4$ properly to form a linear relationship between his and the opponent's payoffs. Since the above unilateral control strategy is realized by setting a determinant to zero, we call this the \emph{zero-determinant strategy under noise} (NZD strategy for short). Note that when there is no noise (i.e., $\tau=1, \varepsilon=0, r=0$), NZD strategy degenerates to the original ZD strategy \cite{Press2012}.


\section{Pinning under Uncertainty}
One specialization of ZD strategies can unilaterally set the opponent's payoff to a deterministic value \cite{Press2012}. Similar strategies were earlier found by Boerlijst, Nowak and Sigmund \cite{Boerlijst1997}. We call such strategies the pinning strategies.
Even in the noisy environments, an NZD strategy can establish a pinning property, although the conditions are more strict. If player X chooses
proper $p_1, p_2, p_3$ and $p_4$, such that $ {\bf{\tilde p}} = \beta {\bf{U}}_{{Y}}  +
\gamma {\bf{1}} $ (set $\alpha=0$), then the following linear equation without player X's payoff involved can be formed：
\begin{eqnarray}\label{pinstrategy}
\beta s_Y  + \gamma  = 0
\end{eqnarray}
The above ${\bf{\tilde p}}$ leads to the following system of linear equations, which depicts the constrains for the pinning strategies under noise:
\begin{eqnarray}
\begin{array}{l}\label{formuc}
\left\{ \begin{array}{l}
 (\tau  + \varepsilon )p_1  + (\varepsilon  + r)p_2   - 1  = \beta R_E + \gamma,  \\
 (\varepsilon  + r)p_1  + (\tau  + \varepsilon )p_2    - 1 = \beta T_E + \gamma,  \\
 (\tau  + \varepsilon )p_3  + (r + \varepsilon )p_4  = \beta S_E + \gamma,  \\
 (r + \varepsilon )p_3  + (\tau  + \varepsilon )p_4  = \beta P_E + \gamma . \\
 \end{array} \right.
\end{array}
\end{eqnarray}

From the first two and the last two equations, we have
$
\beta  = \frac{{\left( {\tau  - r} \right)\left( {p_1  - p_2 }
\right)}}{{R_E - T_E}}
$
and $
\gamma  = p_1  - 1 + \beta\cdot  \frac{{\left( {r + \varepsilon } \right)T_E - \left(
{\tau  + \varepsilon } \right)R_E}}{{\tau  - r}} $, respectively.
There are six variables ($p_1,p_2,p_3,p_4,\beta$ and $\gamma$) in four equations, so we have only two independent free variables. Let $p_1$ and $p_4$ be these two variables, then $p_2$ and $p_3$ can be rewritten as
\begin{eqnarray}
\begin{array}{l}
{p_2} = \frac{1}{A} \cdot {p_1}\left[ {\left( {\tau  + \varepsilon } \right){T_E} + \left( {\varepsilon  + r} \right)\left( {{S_E} - {R_E}} \right) - \left( {\tau  + r} \right){P_E}} \right]\\
{\kern 1pt} {\kern 1pt} {\kern 1pt} {\kern 1pt} {\kern 1pt} {\kern 1pt} {\kern 1pt} {\kern 1pt} {\kern 1pt} {\kern 1pt} {\kern 1pt} {\kern 1pt} {\kern 1pt} {\kern 1pt} {\kern 1pt} {\kern 1pt} {\kern 1pt} {\kern 1pt} {\kern 1pt} {\kern 1pt} {\kern 1pt} {\kern 1pt} {\kern 1pt} {\kern 1pt} {\kern 1pt}  - \frac{1}{A} \cdot \left( {1 + {p_4}} \right)\left( {{T_E} - {R_E}} \right),\\
{p_3} = \frac{1}{A} \cdot {p_4}\left[ {\left( {\tau  + \varepsilon } \right)\left( {{R_E} - {S_E}} \right) + \left( {\varepsilon  + r} \right)\left( {{P_E} - {T_E}} \right)} \right]\\
{\kern 1pt} {\kern 1pt} {\kern 1pt} {\kern 1pt} {\kern 1pt} {\kern 1pt} {\kern 1pt} {\kern 1pt} {\kern 1pt} {\kern 1pt} {\kern 1pt} {\kern 1pt} {\kern 1pt} {\kern 1pt} {\kern 1pt} {\kern 1pt} {\kern 1pt} {\kern 1pt} {\kern 1pt} {\kern 1pt} {\kern 1pt} {\kern 1pt} {\kern 1pt} {\kern 1pt}  - \frac{1}{A} \cdot \left( {1 - {p_1}} \right)\left[ {\left( {\tau  + \varepsilon } \right){P_E} - \left( {\varepsilon  + r} \right){S_E}} \right],
\end{array}
\end{eqnarray}
where $A = \left( {\tau  + \varepsilon } \right)\left( {{R_E} - {P_E}} \right) + \left( {\varepsilon  + r} \right)\left( {{S_E} - {T_E}} \right)$. Representing both $\beta$ and $\gamma$ by $p_1$ and $p_4$ and substituting them back into Eq. (\ref{pinstrategy})
\begin{figure*}
  \subfigure{\label{fig:gull}\includegraphics[width=0.3\textwidth]{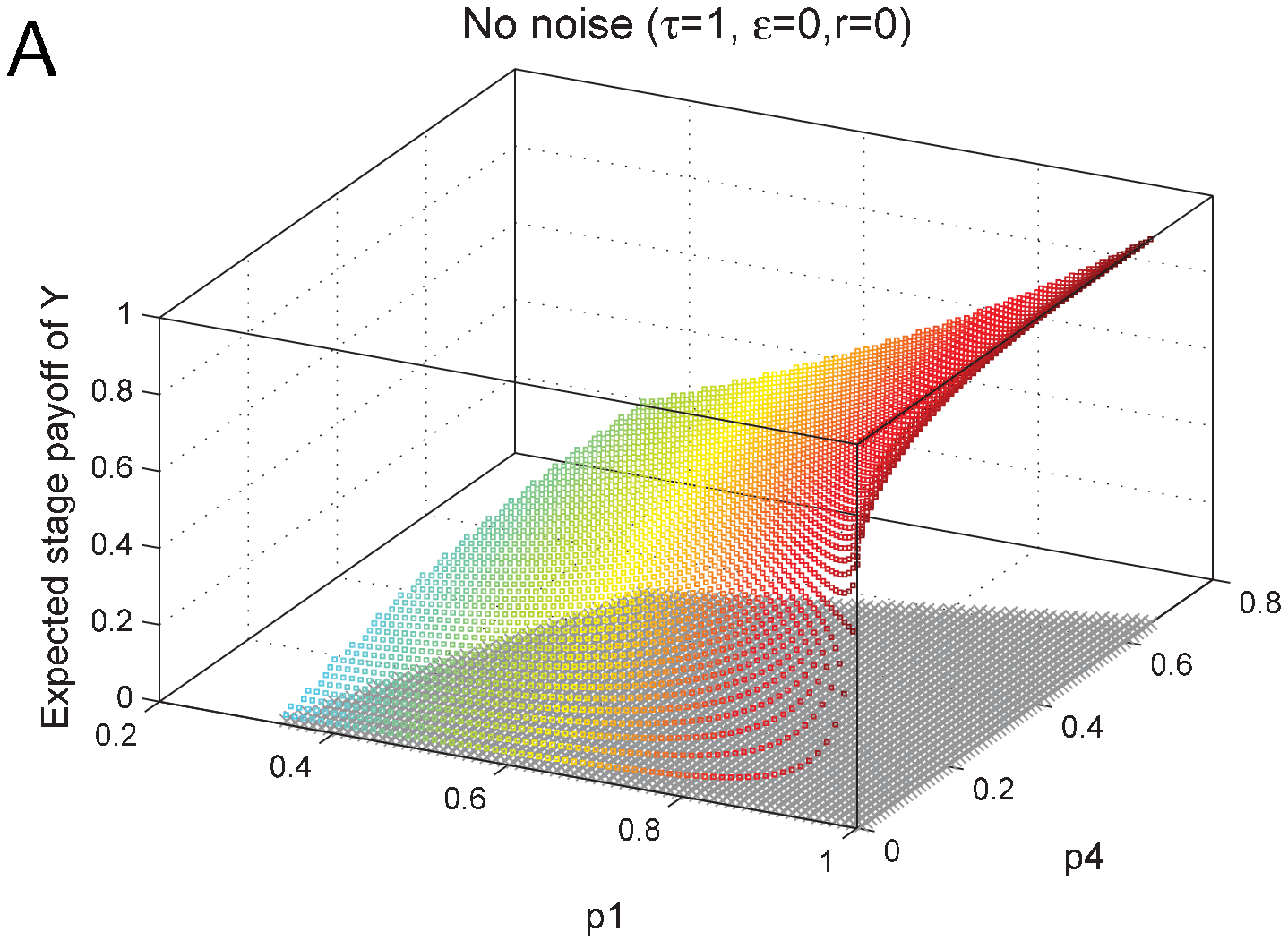}}
  \subfigure{\label{fig:gull}\includegraphics[width=0.3\textwidth]{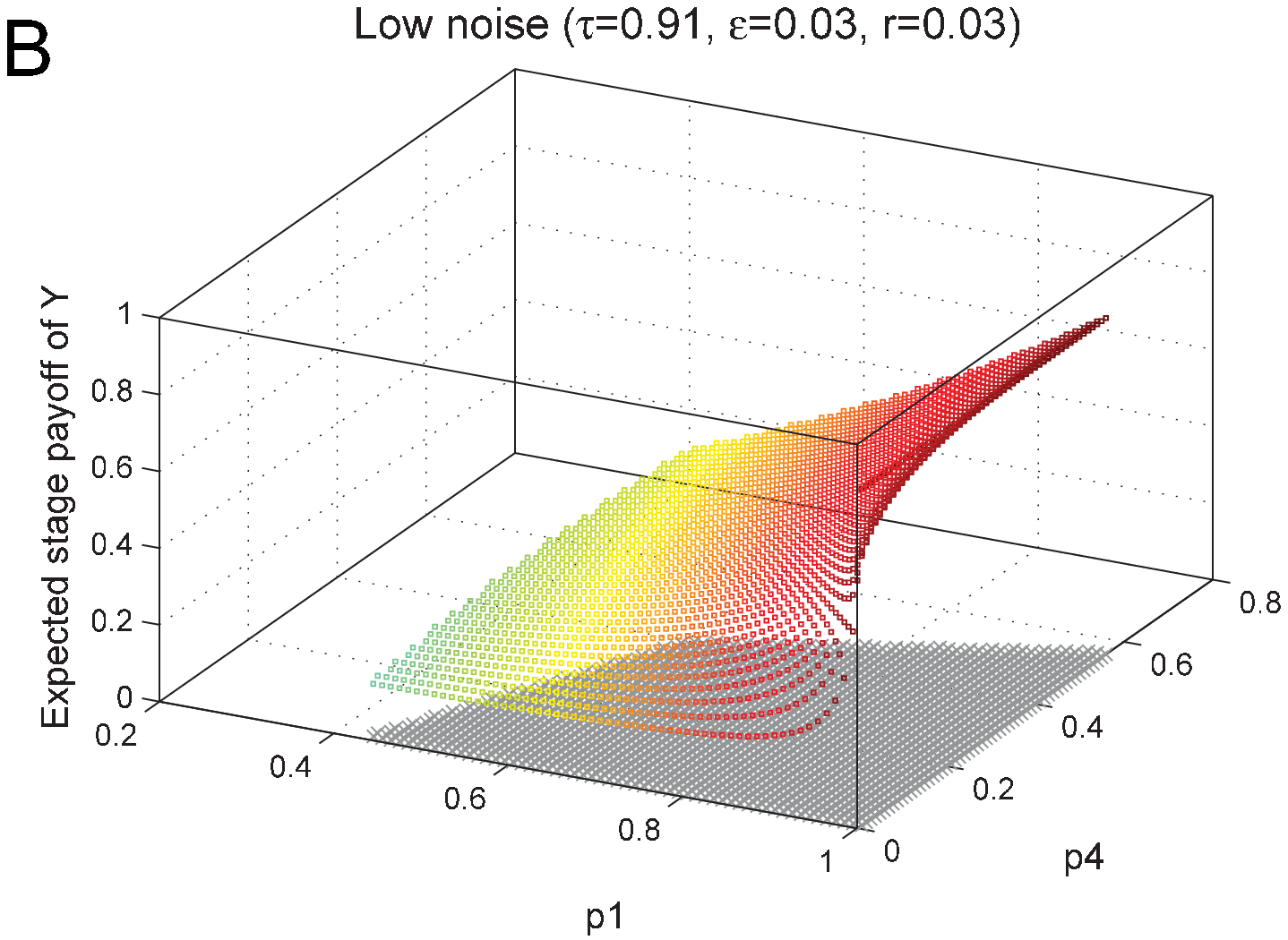}}
  \subfigure{\label{fig:gull}\includegraphics[width=0.3\textwidth]{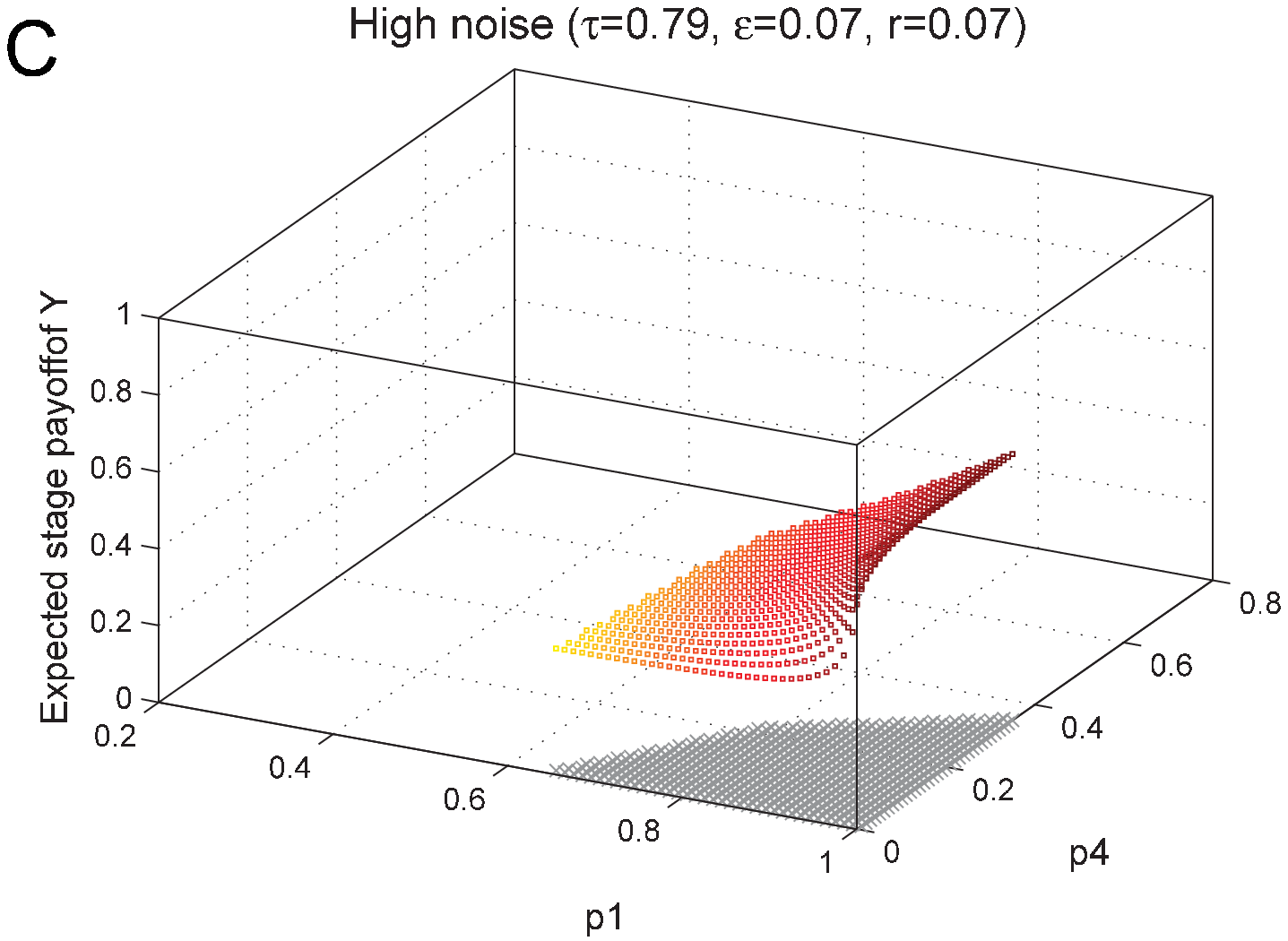}}
  \caption{Feasible region of pinning strategies and the corresponding pinned payoffs of player Y, under different noises. In each sub-figure, the shaded area on the $p_1-p_4$ plane illustrates the feasible region of pinning strategies. The corresponding pinned payoffs are shown as points on the colored surface. The stage game payoffs are calculated by using $G=0.5$ and $L=0.5$, thus realized stage payoffs are $u_i(C,g)=1, u_i(C,b)=-0.5, u_i(D,g)=1.5$ and $u_i(D,b)=0$. The feasible region of pinning strategies as well as the range of pinned payoffs shrink as the noise strength increases. In (A), the game has no noise, thus the expected stage payoffs are $R_E=1, S_E=-0.5, T_E=1.5$ and $P_E=0$. In (B) the game is with a low noise and the expected stage payoffs are $R_E=0.91, S_E=-0.41, T_E=1.4$ and $P_E=0.09$. In (C) there is a high noise and the expected stage payoffs are $R_E=0.79, S_E=-0.29, T_E=1.29$ and $P_E=0.21$.
%
   }
  \label{figure1:feasible_region_and_correponding_payoffs_of_pinning_strategies}
\end{figure*}, we finally get the opponent's payoff, as
\begin{eqnarray}\label{syfinal}
\begin{array}{l}
{s_Y} = \frac{1}{B} \cdot \left( {1 - {p_1}} \right)\left[ {\left( {\tau  + \varepsilon } \right){P_E} - \left( {\varepsilon  + r} \right){S_E}} \right]\\
{\kern 1pt} {\kern 1pt} {\kern 1pt} {\kern 1pt} {\kern 1pt} {\kern 1pt} {\kern 1pt} {\kern 1pt} {\kern 1pt} {\kern 1pt} {\kern 1pt} {\kern 1pt} {\kern 1pt} {\kern 1pt} {\kern 1pt} {\kern 1pt} {\kern 1pt} {\kern 1pt} {\kern 1pt} {\kern 1pt} {\kern 1pt} {\kern 1pt}  + \frac{1}{B} \cdot {p_4}\left[ {\left( {\tau  + \varepsilon } \right){R_E} - \left( {\varepsilon  + r} \right){T_E}} \right],
\end{array}
\end{eqnarray}
where $B = \left( {1 - {p_1} + {p_4}} \right)\left( {\tau  - r} \right)$. It is worth noting that, besides the noise distribution, $s_Y$ is only determined by two components in X's strategic vector, which are $p_1$ and $p_4$. By inspecting the payoff of Y, we found that in the perfect environment ($\tau =1, \varepsilon=r=0$), player Y's payoff degenerates to $
s_Y  = \frac{{\left( {1 - p_1 } \right)P_E + p_4 R_E}}{{\left( {1 - p_1
} \right) + p_4 }}
$.

From Eqs. (\ref{formuc}), the only constrain for the existence of pinning strategies is the probabilistic constrain for $p_1, p_2, p_3$ and $p_4$ (i.e., $0\le p_i\le 1$). We numerically checked the feasible region and the corresponding pinned payoffs of Y, with noise strength ranging from no noise to very strong noise. Since $p_2$ and $p_3$ can be represented by $p_1$ and $p_4$, we only show the feasible region strategies in $p_1-p_4$ plane. As shown in Figure \ref{figure1:feasible_region_and_correponding_payoffs_of_pinning_strategies}(a), the pinned payoff under the perfect environment arches across whole expected payoff space, ranging from $P_E$ to $R_E$. However, as the noise being introduced, on the one hand, the feasible region for pinning strategies shrinks, which indicates the noise brings additional constrains for establishing NZD strategies. On the other hand, the range of the pinned payoff also narrows, showing that the NZD player's power of payoff control will be weakened by the noise. In Figure \ref{figure1:feasible_region_and_correponding_payoffs_of_pinning_strategies}(b), when a weak noise is introduced, the minimum pinned payoff is higher than $P_E$ and the maximum pinned payoff is lower than $R_E$, and as shown in Figure \ref{figure1:feasible_region_and_correponding_payoffs_of_pinning_strategies}(c), with the noise strength, the range of the pinned payoff continuously reduces to a very narrow one.

\section{Extortion under Uncertainty}
An NZD strategy in Eqs. (\ref{equationset}) can be equivalently rewritten as
\begin{eqnarray}\label{general_form_ZD}
\widetilde {\bf{p}} = \varphi \left[ {\left( {{{\bf{U}}_X} - l{\bf{1}} } \right) - \chi \left( {{{\bf{U}}_Y} - l{\bf{1}}} \right)} \right],
\end{eqnarray}
where $\varphi$, $\chi$ and $l$ are free parameters. The only usage of $\varphi$ is to ensure the probabilities to locate in $[0,1]$. It is worth noting that if $l \le P_E$, the probability constrains cannot be satisfied and NZD strategies do not exist. Thus we only need to investigate different cases when $l \ge P_E$. In the case (i) $\chi \to \infty$, $\bf p$ is a pinning strategy. In the case (ii) $\chi > 1$ and $l\ge P_E$, player X can ensure that, when player Y tries to increase his payoff, he will increase X's even more, and X's increase of payoff exceeds that of Y by a fixed percentage $\chi$. In addition, Y can only maximize his payoff by fully cooperating $({\bf q}={\bf 1})$. Therefore, if player X chooses a $\bf p$ with $\chi > 1$, then X can always extort Y since Y's effort will benefit X more than himself. In the case (iii) $\chi > 1$ and $l=P_E$, player X not only ensures his payoff increment is $\chi$-fold of Y's, but also guarantees that his absolute payoff is always higher than Y's, and consequently dominates in the game. Therefore, we distinguish the second and the third cases, and call the former \emph{weak extortion} strategy and the later \emph{strong extortion} strategy. It is worth noting that, a strong extortion strategy is the most stringent case of the weak extortion strategies. Essentially, the strength of extortion is quantitatively affected by the parameter $l$, which can be seen as the \emph{baseline} of extortion.


%



Although the strong extortion strategies are found widely existing in games without noise \cite{Press2012}, we prove that in noisy repeated games, the strong extortion strategies do not exist. To enforce a strong extortion strategy, according to Eq. (\ref{mischivous}), the following equation set is required to be satisfied when $l=P_E$.
{\small
\begin{eqnarray}
\begin{array}{l}\label{special_case_of_pinning}
\left\{ \begin{array}{l}
  (\tau  + \varepsilon ){p_1} + (\varepsilon  + r){p_2} - 1 = \varphi \left[ {\left( {{R_E} - {l}} \right) - \chi \left( {{R_E} - {l}} \right)} \right],\\
  (\varepsilon  + r){p_1} + (\tau  + \varepsilon ){p_2} - 1 = \varphi \left[ {\left( {{S_E} - {l}} \right) - \chi \left( {{T_E} - {l}} \right)} \right],\\
(\tau  + \varepsilon ){p_3} + (r + \varepsilon ){p_4} = \varphi \left[ {\left( {{T_E} - {l}} \right) - \chi \left( {{S_E} - {l}} \right)} \right],\\
(r + \varepsilon ){p_3} + (\tau  + \varepsilon ){p_4} = \varphi \left[ {\left( {{P_E} - {l}} \right) - \chi \left( {{P_E} - {l}} \right)} \right].
\end{array} \right.
\end{array}
\end{eqnarray}
}However, when $l=P_E$, the third and the fourth equations can not be satisfied simultaneously.
Intuitively, the missing of strong extortion strategy in noisy repeated games is due to the reason that, the errors introduce stochasticity and uncertainty into the payoffs, and consequently has an negative impact on the accuracy of player X's payoff-based strategy setting. Therefore, the NZD player faces a fundamental tradeoff between the payoff control ability and the payoff dominance. Such a tradeoff is similar to the relationship between the risk dominance and payoff dominance, which has been discussed in pioneering works by Harsanyi and Selten \cite{Harsanyi1998}. Thus in a noisy environment, to regain the payoff control ability, the extortioner needs to relax the extortion baseline from $P_E$ to $P_E+\Delta$, which, on the contrary, increases the risk for him to loss in payoff. We represent the weak extortion strategy as $(\chi,\Delta)$-extortion strategy, where $\chi$ defines the extortion rate while $\Delta=l-P_E$ defines the distance between the weak and strong extortion strategies that can be considered as the \emph{generosity} \cite{Stewart2013}. When $\Delta$ is small, it is still very likely (though not necessarily) for player X to always get higher payoffs than player Y, however, it will be difficult for her to establish an extortion on player Y's payoff. A larger $\Delta$ indicates that player X offers more opportunity for the opponent to win in payoff, but correspondingly obtains higher possibility for himself to control the opponent's payoff. Therefore, in order to realize a payoff control while reducing the risk of losing, it is of great importance for NZD player to design his strategy with a proper extortion ratio $\chi$ and a sufficiently small distance $\Delta$.
 \begin{figure}
  \subfigure{\label{fig:gull}\includegraphics[width=0.23\textwidth]{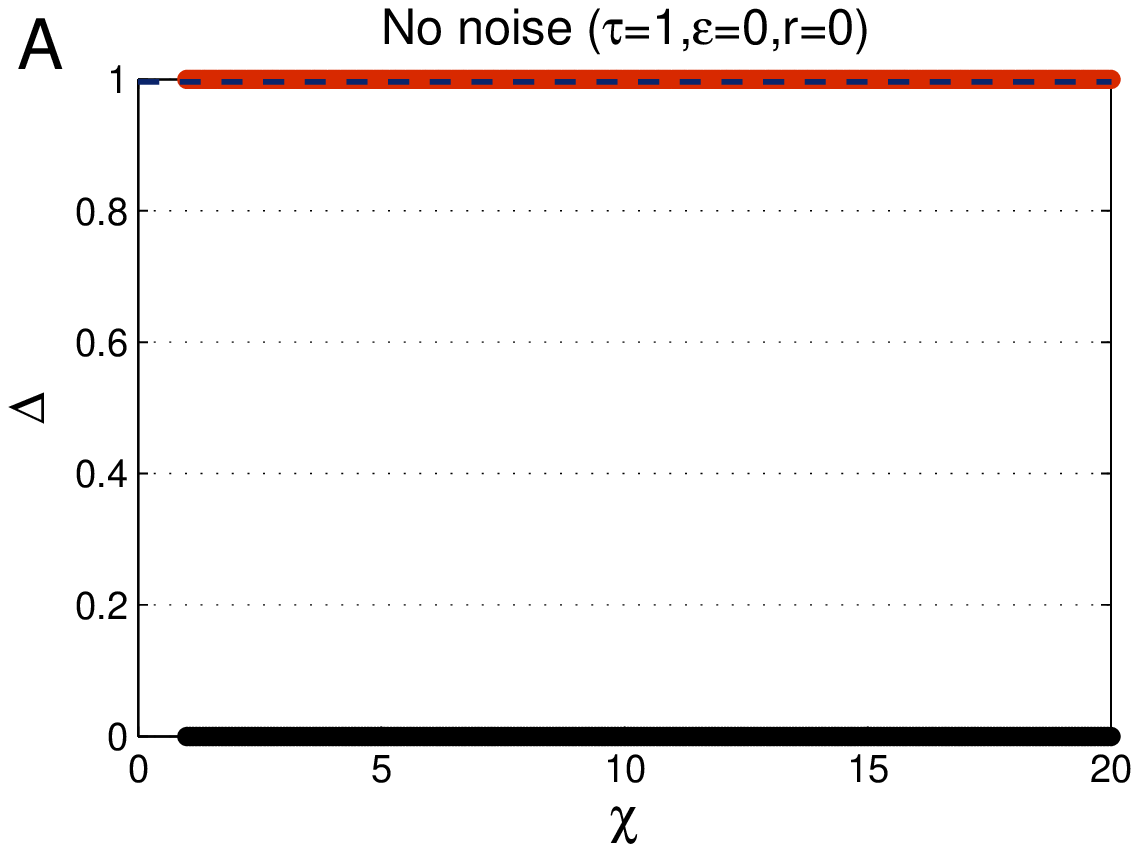}}
  \subfigure{\label{fig:gull}\includegraphics[width=0.23\textwidth]{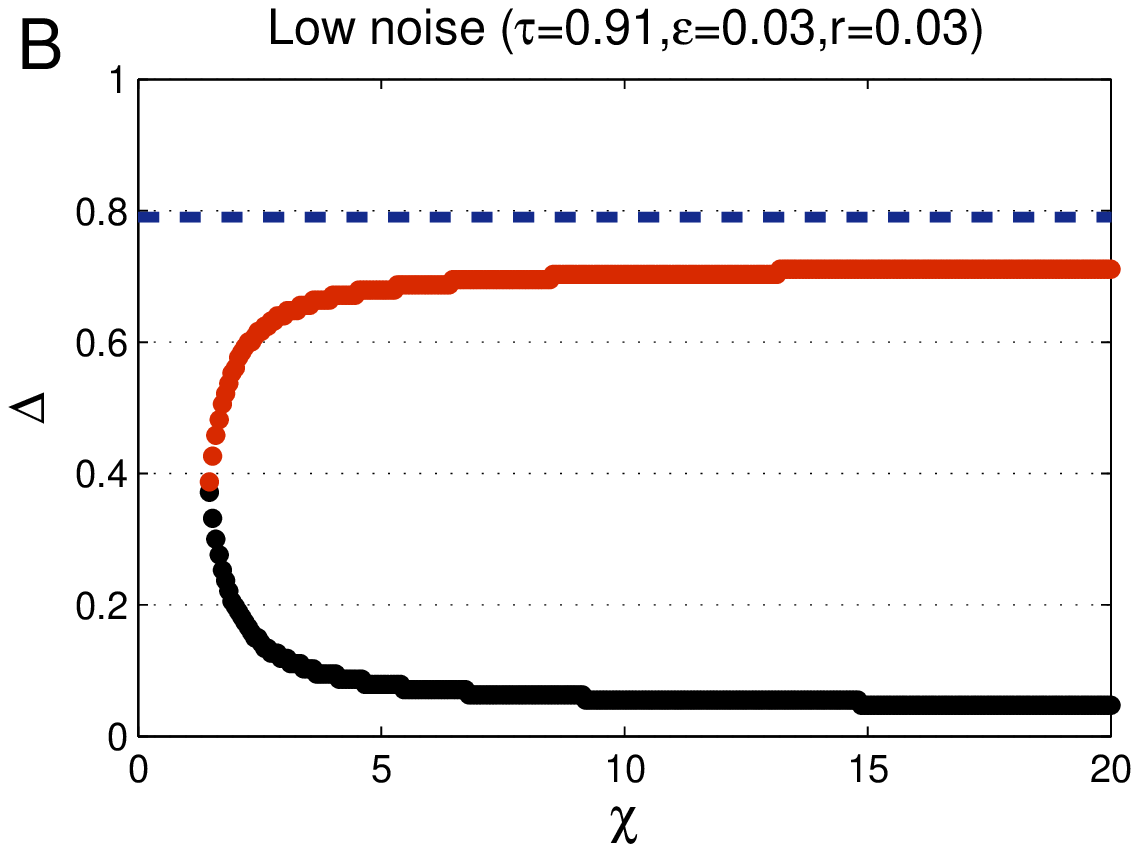}}
    \subfigure{\label{fig:gull}\includegraphics[width=0.23\textwidth]{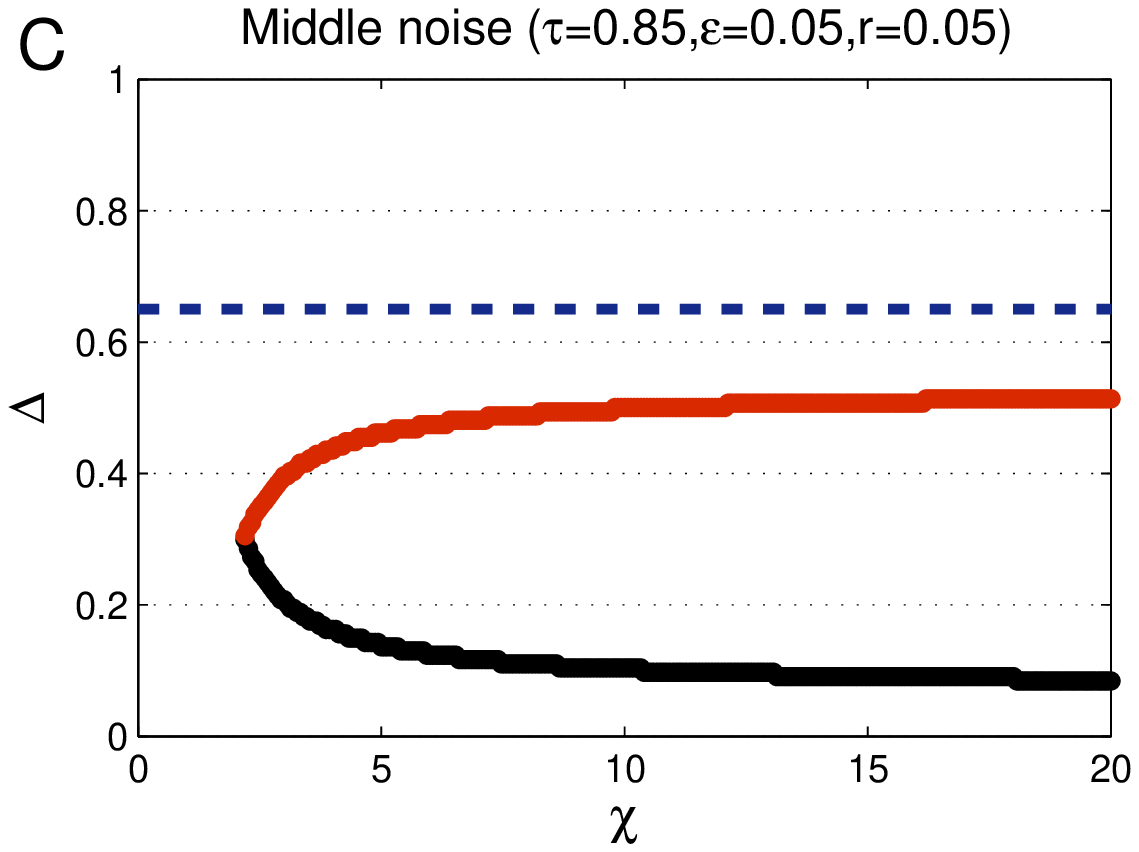}}
      \subfigure{\label{fig:gull}\includegraphics[width=0.23\textwidth]{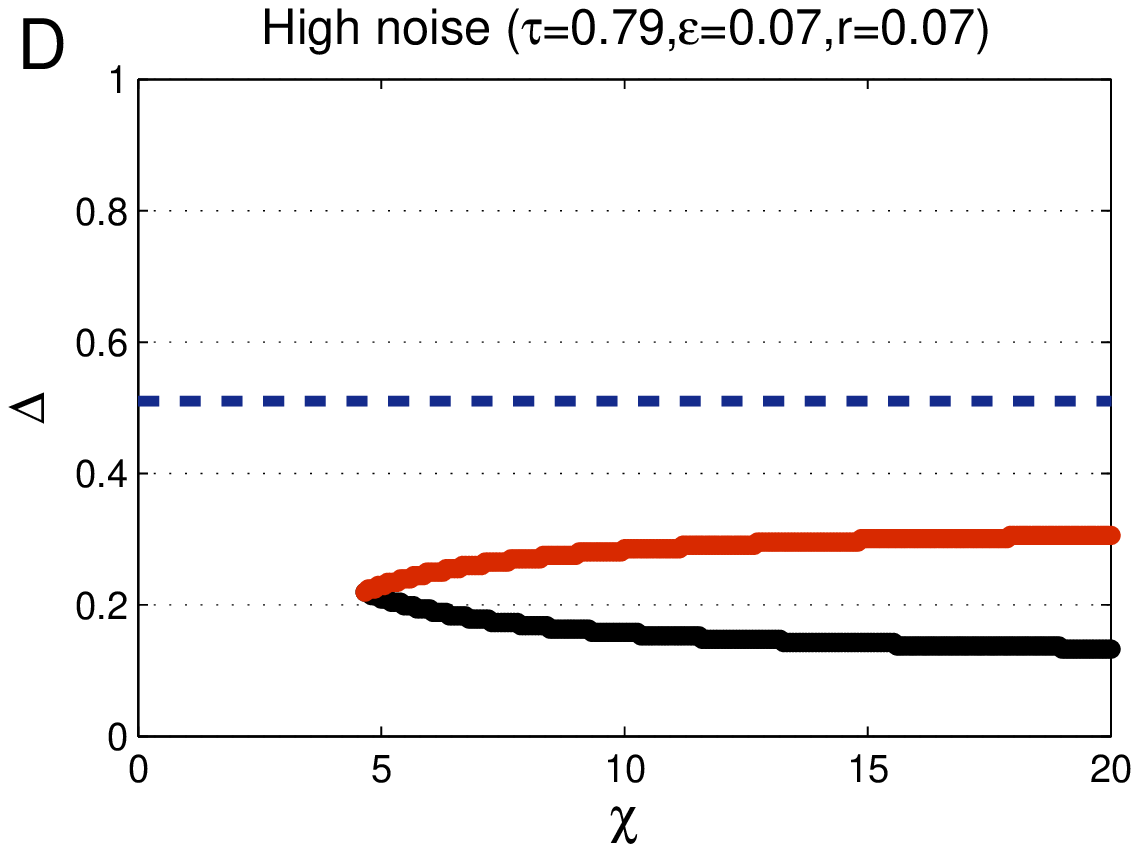}}
  \caption{Feasible regions for weak extortion strategies under different noise strengths. The black curves and red curves depict the lower bounds and upper bounds of $\Delta$ versus $\chi$, respectively. The blue dash lines show the values of $R_E-P_E$. In all sub-figures, player X's realized payoff is set as ${u_X}\left( {C,g} \right)=1$, ${u_X}\left( {C,b} \right)=-0.5$, ${u_X}\left( {D,g} \right)=2$ and ${u_X}\left( {D,b} \right)=0$. The expected stage payoffs $R_E$, $S_E$, $T_E$ and $P_E$ are calculated by Eq. (\ref{expected_stage_payoff}).  For the noise-free case (A), the lower bound of $\Delta$ is always $0$ and the upper bound of $\Delta$  is always $1$, indicating that NZD strategies always exist for any $\chi$. In the low noise case (B), the weak extortion strategies with small $\chi$ ($\chi<1.78$) do not exist, and the feasible range of $\Delta$ becomes larger as the increase of $\chi$ after it exceeds $1.78$. The lower bound approaches a value greater than $0$ while the upper bound approaches a value smaller than $0.79$. In (C), there is a middle noise, and in (D), there is a very strong noise. Comparing (A), (B), (C) and (D), it is found that the feasible region of weak extortion strategies dramatically shrinks with the increase of noise strength. }
  \label{feasible_region_extortion}
\end{figure}

According to the analysis above, to get a weak extortion strategy under noise, the following vector equation is required:
{\small
\begin{eqnarray}\label{delta_extortion}
\widetilde {\bf{p}} = \varphi \left[ {\left( {{{\bf{U}}_X} - (P_E+\Delta){\bf{1}} } \right) - \chi \left( {{{\bf{U}}_Y} - (P_E+\Delta){\bf{1}}} \right)} \right],
\end{eqnarray}
}
which can be expanded to:
\begin{eqnarray}
\begin{array}{l}
{p_1} =   1-\varphi\frac{1}{\tau-r}\left[ F_1-\chi F_2 \right]+\varphi \left( {\chi  - 1} \right)\Delta,\\
\\
{p_2} =   1+\varphi\frac{1}{\tau-r}\left[ J_1-\chi J_2 \right]+\varphi \left( {\chi  - 1} \right)\Delta,\\
\\
{p_3} =  \varphi\frac{ \left( {\tau  + \varepsilon } \right) }{{\tau-r }}\left[ {T_E - P_E - \chi \left( {S_E - P_E} \right)} \right]{\kern 1pt}  + \varphi \left( {\chi  - 1} \right)\Delta, \\
\\
{p_4} = -\varphi \frac{ \left( {\varepsilon  + r} \right)}{{\tau-r}}\left[ {T_E - P_E - \chi \left( {S_E - P_E} \right)} \right] + \varphi \left( {\chi  - 1} \right)\Delta,
\end{array}
\end{eqnarray}
where $F_1=\left( {\tau  + \varepsilon } \right){R_E} -\left( {\varepsilon  + r} \right){S_E} + \left( {r - \tau } \right){P_E}$, $F_2= {\left( {\tau  + \varepsilon } \right){R_E} - \left( {\varepsilon  + r} \right){T_E} + \left( {r - \tau } \right){P_E}} $, $J_1=\left( {\varepsilon  + r} \right){R_E} - \left( {\tau  + \varepsilon } \right){S_E} + \left( {\tau  - r} \right){P_E}$ and $J_2= {\left( {\varepsilon  + r} \right){R_E} - \left( {\tau  + \varepsilon } \right){T_E} + \left( {\tau  - r} \right){P_E}}  $.
As shown in Figure \ref{feasible_region_extortion}, we numerically checked the feasible region of weak extortion strategies by exploring the whole space of $\Delta$ versus different extortion ratio $\chi$. One can see that the distance
$\Delta$ has both lower bound and upper bound, with the former positively correlated with the noise strength and the latter negatively correlated with the noise strength.
Combining these two effects, the feasible range of $\Delta$ shrinks while the noise becomes stronger. In addition, the increasing of lower bound suggests that the NZD player should relax its extortion baseline $l$ and move it farther from $P_E$ as the noises strength increases.

When player X adopts a weak extortion strategy, the payoffs of
players X and Y follow the following linear relationship:
\begin{eqnarray}\label{sx_sy}
s_X  - \left( {P_E + \Delta } \right) = \chi \left[ {s_Y  - \left( {P_E
+ \Delta } \right)} \right].
\end{eqnarray}
Since in the Prisoner's Dilemma, $T_E>R_E>P_E>S_E$, X's payoffs when Y chooses action $C$ ($T_E$ or $R_E$) are always larger than his payoffs when Y chooses action $D$ ($P_E$ or $S_E$). The same
result holds when player Y mixes his action. Thus whatever
strategy X takes, its expected payoff $s_X$ will be maximized when
Y fully cooperates ($\bf q=1$). When X takes weak extortion strategy, since $s_X$ and $s_Y$ follow a linear relationship, $s_Y$ will also be
maximized when $s_X$ reaches its maximum. Therefore, both $s_X$ and
$s_Y$ are maximized when Y fully cooperates. Substituting $q_1=q_2=q_3=q_4=1$
into $\det ({\tilde{\bf{M}}})$, the determinant becomes
\begin{eqnarray}
\begin{array}{l}\label{detmprime_reformed}
\det ( {\tilde{\bf{M}}} )=\det \left( {{\bf{p}},{\bf{1}},{\bf U}_X }
\right) \\= \left| {\begin{array}{*{20}c}
   {1 - \left( {\tau  + \varepsilon } \right)p_1  + \left( {\varepsilon  + r} \right)p_2 } & {\kern 5pt}0 & {\kern 5pt}0 & {\kern 5pt}R_E  \\
   {\left( {\varepsilon  + r} \right)p_1  + \left( {\tau  + \varepsilon } \right)p_2 } & {\kern 5pt}{ - 1} & {\kern 5pt}1 & {\kern 5pt}S_E  \\
   {\left( {\tau  + \varepsilon } \right)p_3  + \left( {\varepsilon  + r} \right)p_4 } & {\kern 5pt}0 & {\kern 5pt}0 & {\kern 5pt}T_E  \\
   {\left( {\varepsilon  + r} \right)p_3  + \left( {\tau  + \varepsilon } \right)p_4 } & {\kern 5pt}0 & {\kern 5pt}1 & {\kern 5pt}P_E  \\
\end{array}} \right|.
\end{array}
\end{eqnarray}
Making Laplace expansion on the fourth column, we have
\begin{eqnarray}
\begin{array}{l}\label{detmprime_reformed_expansion}
\begin{array}{l}
\det \left( {{\bf{p}},{\bf{1}},{\bf U}_X } \right)
 \\=  - R_E \cdot \left| {\begin{array}{*{20}c}
   {\left( {\varepsilon  + r} \right)p_1  + \left( {\tau  + \varepsilon } \right)p_2 } & {\kern 5pt}{ - 1} & {\kern 5pt}1  \\
   {\left( {\tau  + \varepsilon } \right)p_3  + \left( {\varepsilon  + r} \right)p_4 } & {\kern 5pt}0 & {\kern 5pt}0  \\
   {\left( {\varepsilon  + r} \right)p_3  + \left( {\tau  + \varepsilon } \right)p_4 } & {\kern 5pt}0 & {\kern 5pt}1  \\
\end{array}} \right|
\\- T_E\cdot\left| {\begin{array}{*{20}c}
   {1 - \left( {\tau  + \varepsilon } \right)p_1  + \left( {\varepsilon  + r} \right)p_2 } & {\kern 5pt}0 & {\kern 5pt}0  \\
   {\left( {\varepsilon  + r} \right)p_1  + \left( {\tau  + \varepsilon } \right)p_2 } & {\kern 5pt}{ - 1} & {\kern 5pt}1  \\
   {\left( {\varepsilon  + r} \right)p_3  + \left( {\tau  + \varepsilon } \right)p_4 } & {\kern 5pt}0 & {\kern 5pt}1  \\
\end{array}} \right|.  \\
 \end{array}
\end{array}
\end{eqnarray}
The normalized payoff for player X is then
$$
s_X  = \frac{{\det \left( {{\bf{p}},{\bf{1}},{\bf{U}}_X }
\right)}}{{\det \left( {{\bf{p}},{\bf{1}},{\bf{1}}} \right)}},
$$which finally leads to
\begin{eqnarray}\label{sx_analytical}
\begin{array}{l}
{s_X} = \frac{1}{C}  \chi \left[ {{R_E}\left( {{T_E} - {S_E}} \right) - {P_E}\left( {{T_E} - {R_E}} \right)} \right]\\
{\kern 1pt} {\kern 1pt} {\kern 1pt} {\kern 1pt} {\kern 1pt} {\kern 1pt} {\kern 1pt} {\kern 1pt} {\kern 1pt} {\kern 1pt} {\kern 1pt} {\kern 1pt} {\kern 1pt}  - \frac{1}{C}  \left( {\chi  - 1} \right) \left( {{T_E} - {R_E}} \right)\Delta + \frac{1}{C} {P_E}\left( {{T_E} - {R_E}} \right),
\end{array}
\end{eqnarray}
and
\begin{eqnarray}\label{sy_analytical}
\begin{array}{l}
 s_Y  = \frac{1}{C}\chi P\left( {R_E - S_E} \right) + \frac{1}{C}\left( {\chi  - 1} \right)\Delta \left( {R_E - S_E} \right) \\
 {\kern 1pt} {\kern 1pt} {\kern 1pt} {\kern 1pt} {\kern 1pt} {\kern 1pt} {\kern 1pt} {\kern 1pt} {\kern 1pt} {\kern 1pt} {\kern 1pt} {\kern 1pt} {\kern 1pt} {\kern 1pt}  + \frac{1}{C}\left[ {S_E\left( {P_E - R_E} \right) + P_E\left( {T_E - R_E} \right)} \right], \\
\end{array}
\end{eqnarray}
where $C = \left( {{T_E} - {R_E}} \right) + \chi \left( {{R_E} - {S_E}} \right)>0$. For instance, if $(R_E,S_E,T_E,P_E)=(3,0,5,1)$, we have
\begin{equation}\label{sx_numerical}
s_X  = \frac{{2 + 13\chi  - 2 \left( {\chi  - 1} \right)\Delta}}{{2
+ 3\chi }},
\end{equation}
and accordingly, the payoff for player Y is
\begin{equation}\label{sy_numerical}
s_Y = \frac{{12 + 3\chi  + 3 \left( {\chi  - 1} \right)\Delta}}{{2
+ 3\chi }}.
\end{equation}

In a word, on the one hand, the extortion strategies are still feasible in noisy environment, which indicates it is still possible for the NZD player to ensure that when the opponent tries to improve his payoff, he will improve the NZD player's even more. And the opponent will maximize his own payoff by fully cooperating, where the NZD player's payoff is also maximized. Thus the NZD player can still enforce a weak extortion on his opponent. However, on the other hand, the uncertainty in the noisy environment has abated the power of extortion, in the sense that the extortioner cannot guarantee his payoff to be always higher than the opponent's and the strong extortion strategies do not exist. The baseline for weak extortion strategies should have a distance to $P_E$, and
the lower bound of the distance has a positive correlation with noise strength. Under a same extortion ratio $\chi$, the payoffs for the extortioner and for the opponent under different noise strengths varies. In Eq. (\ref{sx_analytical}) we can see
$s_X$ may decline as noisy strength increases. On the contrary, in Eq. (\ref{sy_analytical}), $s_Y$ may increase as noisy strength increases.
Therefore under a certain noise strength (which results in a reasonably large distance), it is possible for $s_Y$ to outperform $s_X$.
These indicate in noisy environments, when an NZD player wishes to extort the opponent and control the payoffs, there rises a risk for her to loss in payoff, especially when the noise is strong. Therefore, in a realistic uncertain world, extorting others has the potential to cause damage to yourself.

\section{Conclusion and Discussion}
The concept of ZD strategy has become a promising framework to explore the long-run relationships. However, out of the laboratory, the existence of noises in the environment elevates the complexity of games and the payoff-oriented ZD strategy selection in such games deserves more concrete analysis.
We established the generalized
form of ZD strategy for noisy games and named it NZD strategy. We identify three specifications of NZD strategies, namely the pinning strategies, strong extortion and weak extortion. We also study the conditions, feasible regions and
corresponding payoffs for these strategies. It is found that NZD strategies have high robustness against noise and widely exist in noisy games with reasonable noise strength, although the noise has negative impact on the existence and performance
of NZD strategies. The noises will expose the NZD player to uncertainty and risk, however, it is still possible for him to set the opponent’s payoff to a fixed value, or to extort the opponent.

The implementation of the NZD strategies relays on the existence of the unique stationary distribution. However, not only the existence of noisy but also some special strategies, may result in bad circumstances such that the regularity of Markov matrix cannot be satisfied, or the Markov process may not converge to a unique stationary distribution. Thus it is essential to analyze the convergency of the Markov process of the game. This is not only important to ZD or NZD strategies, but also a key problem for other topics in repeated games. When multiple stationary distribution exists, the Markov process may have multiple converging states, which belongs to different communicating classes. In this case, the expected payoff of each player is strongly affected by the initial state of the game. We conjecture that, in a game with multiple stationary distribution, a generalized NZD strategies whose expected payoff is engaged with initial distribution, may still exist. Moreover, the speed for the Markov process to converge is a key factor for the NZD player. The second-largest eigenvalue of a Markov transition matrix is a convenient factor to determine which strategy of the NZD player may lead the game to converge faster. Although the converging speed is not unilaterally determined by the NZD player, he can at least secure himself with a maximized lower boundary of the converging speed.

Furthermore, the original ZD strategies are not necessarily promoting cooperations, since the Markov process does not surely converge to a joint state $CC$. When the repeated game is played in an imperfect environment, this becomes even more severe. The \emph{generous strategies} \cite{Stewart2013} not only guarantee a linear relationship between two players' payoffs, but also ensure that the mutual cooperation payoff is the maximum payoff to both the ZD player and the opponent. Generosity comes at a cost, but it finally encourages everybody to cooperate. Although the generous strategies are proved to be very robust in the perfect environment, whether it exists and how it performs in the noisy environment still need investigation. In particular, how to provide a strategy that makes the game always converge to the mutual cooperation state, even when the noisy have disturbance on the mutual cooperation? Actually, this topic is strongly related to the equilibrium analysis in repeated games with private monitoring, which is the one of the most well-known long-standing open problems in game theory research \cite{Kandori2002}. The framework of NZD strategies may potentially provides us with another possible direction to tackle this issue.



\section*{acknowledgement}
This work was partially supported by the National Natural Science Foundation of China under Grant Nos. $61473060$ and $11222543$.


\begin{thebibliography}{99}
\bibitem{M&LBOOK}
J. Mailath and L. Samuelson,  \emph{Repeated Games and Reputation} (Oxford University Press, 2006).
\bibitem{Press2012}
W. H. Press and F. J. Dyson, Proc. Acad. Natl. Sci. U.S.A. {\bf 109}, 10409 (2012).
\bibitem{Stewart2012}
A. J. Stewart and J. B. Plotkin, Proc. Acad. Natl. Sci. U.S.A. {\bf 109}, 10134 (2012).
\bibitem{Hayes2013}
B. Hayes, American Scientist {\bf 101}, 422 (2013).
\bibitem{Hao2014}
D. Hao, Z. Rong, and T. Zhou, Chin. Phys. B {\bf 23}, 078905, (2014).

\bibitem{Roemheld2013}
L. Roemheld, arXiv: 1308.2576 (2013).

\bibitem{Akin2012}
E.  Akin,  arXiv: 1211.0969 (2012).

\bibitem{Stewart2013}
 A. J. Stewart and J. B Plotkin, Proc. Acad. Natl. Sci. U.S.A. {\bf 110}, 15348 (2013).
 \bibitem{Hilbe2013Plos}
C. Hilbe, M. A. Nowak, and A. Traulsen, PLoS ONE {\bf 8}, e77886 (2013).
\bibitem{Chen2014}
J. Chen and A. Zinger, J. Theor. Biol. {\bf 357}, 46 (2014).
\bibitem{Pan2014}
L. Pan, D. Hao, Z. Rong, and T. Zhou, arXiv: 1402.3542 (2014).

\bibitem{Hilbe2014}
C. Hilbe, B. Wu,  A. Traulsen, and M. A. Nowak, Proc. Acad. Natl. Sci. U.S.A. {\bf 111}, 16425 (2014).

\bibitem{Hilbe2014-2}
C. Hilbe, T. Rohl, and M. Milinski, Nature Commun. {\bf 5}, 3976 (2014).





\bibitem{Adami2013}
C. Adami and A. Hintze, Nature Commun. {\bf 4}, 2193 (2013).
\bibitem{Hilbe2013}
C. Hilbe, M. A. Nowak, and K. Sigmund, Proc. Acad. Natl. Sci. U.S.A. {\bf 110}, 6913 (2013).

\bibitem{Szolnoki2014}
A. Szolnoki and M. Perc, Phys. Rev. E {\bf 89}, 022804 (2014).

\bibitem{Wu2014}
Z. X. Wu and Z. Rong, Phys. Rev. E {\bf 90}, 062102 (2014).


\bibitem{Szolnoki2014-2}
A. Szolnoki and M. Perc, Sci. Rep. {\bf 4}, 5496 (2014).


\bibitem{Kandori2002}
M. Kandori, J. Econ. Theor. {\bf 102}, 1 (2002).
\bibitem{Nowak1995}
M. A. Nowak, K. Sigmund, and E. El-Sedy, J. Math. Biol. {\bf 33}, 703 (1995).
\bibitem{Fudenberg2012}
D. Fudenberg, G. R. David, and D. Anna, Am. Econ. Rev. {\bf 102}, 720 (2012).
\bibitem{Fudenberg1990}
D. Fudenberg and E. Maskin, Am. Econ. Rev. {\bf 80}, 274 (1990).
\bibitem{Sekiguchi1997}
T. Sekiguchi, J. Econ. Theor. {\bf 76}, 345 (1997).
\bibitem{Barlo2009}
M. Barlo, C. Guilherme, and S. Hamid, J. Econ. Theor. {\bf 144}, 312 (2009).
\bibitem{Mailath2002}
G. J. Mailath and M. Stephen,  J. Econ. Theor. {\bf 102}, 189 (2002).
\bibitem{Mailath2011}
G. J. Mailath and O. Wojciech, Gam. Econ. Behav. {\bf 71}, 174 (2011).
\bibitem{Harsanyi1998}
J. C. Harsanyi and R. Selten, \emph{A General Theory of Equilibrium Selection in Games} (MIT Press, 1988).

\bibitem{Murdock1962}
B. B. Murdock, J. Exper. Psy. {\bf 64}, 482 (1962).
\bibitem{Boerlijst1997}
M. C. Boerlijst, M. A. Nowak, and K. Sigmund, Amer. Math. Mon. {\bf 104}, 303 (1997).
\end{thebibliography}
\end{document}